\documentclass{article}[12pt]
\usepackage{epsfig}
\begin{document}
{\Huge{Numerical calculation of the energy relative fluctuation for a system in contact with a finite heat bath.\\}}

\vspace{0.5cm}
{\Large{F. Q. Potiguar\footnote{email: potiguar@fisica.ufc.br}, U. M. S. Costa\\}}
{\large{\em{Universidade Federal do Cear\'a, Departamento de F\'\i sica, Campus do Pici, 60455-760, Fortaleza, Cear\'a, Brasil\\}}}

\begin{center}
\section*{Abstract}
\end{center}
We use a scheme of separation of degrees of freedom for a system in order to produce two systems with finite number of degrees of freedom. Our intent is to measure the energy square relative fluctuation (SRF) of the observable part through the simulation of two simple examples of composed systems, the simple harmonic oscillator, and the chain of quartic oscillators. We want to test the result found previously by us through the finite heat bath canonical ensemble (cond-mat/0210525), which is an application of Tsallis' statistics. We see that the results obtained here are in very good agreement with the theoretical predicted values. This suggests that this kind of finite systems are ergodic, and that they do not provide ``bad'' statistics.\\

\vspace{0.5cm}
PACs: 05.20.-y, 05.70.-a


\newpage
\section{Introduction}
\label{section1}
\noindent
Tsallis' statistics, first proposed 14 years ago \cite{Tsa88}, is a field which attracts great attention nowadays. It has been applied to a variety of fields, among which are included: fully developed turbulence \cite{Bec02}, anomalous diffusion \cite{Levy}, and the Nos\'e-Hoover thermostat \cite{Nose}.\\
Certainly, many proposals to explain its physical meaning, in particular the interpretation of the generalization parameter $q$, have risen. Some of them \cite{q-int} suggest that $q$ measures the fluctuation of a quantity pertaining the phenomenom in which the interpretation was proposed.\\
We follow the path initially proposed by Plastino and Plastino \cite{PlasPlas94}, and Almeida \cite{Alm01}. This approach is based on the idea that Tsallis' power law canonical distribution is the probability function of a system which is in thermal equilibrium with a heat bath which has a {\em finite number of degrees of freedom}, and the quantity $\frac{q_i}{q_i-1}$ is the exponent of the energy in the accessible phase space volume of the $i$-th system which is proportional to the number of degrees of freedom of this system. It may either be positive or negative.\\
It should be noticed that in this formalism the distribution is obtained in the first place (instead of postulating an entropy functional and then finding the canonical form by optimization of the former) from the system's phase space geometry.\\
Of course, this is only one case where Tsallis' statistics can be said to ``apply'', since the generalization parameter $q$, firstly free of interpretation, provides a very convenient way of describing a wide class of phenomena. Henceforth, we will refer to our framework as the finite bath canonical ensemble, but still use the notation introduced previously. Any quantities labeled by $0$, $1$, and $2$ belong to the composite (observable+bath), observable, and heat bath system.\\
Many theoretical developments were achieved along this line \cite{Pros93,Alm02-01,Alm02-02,PoC02-01}. Also, a few simulations were performed in order to test the present approach \cite{Adib02-01,Adib02-02}. A detailed description of the finite heat bath canonical ensemble may be found within these references.\\
We propose here to simulate two simple systems in order to measure the square relative fluctuation (SRF) of a system. We want to test the result found in \cite{PoC02}. We do this by using a procedure in which we separate the degrees of freedom of a classical system, which is not is thermal contact with any heat bath, and define one of them as the observable system, and the other as the bath. This was primarly used in \cite{Adib02-01,Adib02-02}.\\
We organize the paper as follows: in section 2, we give a brief explanation of the relative fluctuation calculation made in \cite{PoC02}. Section 3 is devoted to introduce the model, the systems, and the simulations' results. In section 4 we discuss the physical meaning of this approach. In section 5 we address our conclusions.\\

\section{The relative fluctuation in the finite bath approach}
\label{section2}
\noindent
In order to calculate the SRF of a random quantity $x$, we need to calculate the following expression:
\begin{equation}
\label{x-relfluct}
\frac{\left<\Delta x^2\right>}{\left<x\right>^2}=\frac{\left<x^2\right>}{\left<x\right>^2}-1.
\end{equation}
The angular brackets denote ensemble averages.\\
In \cite{PoC02}, the authors calculated equation (\ref{x-relfluct}) for a system's energy $E_1$. This energy is distributed according to the following power law canonical distribution:
\begin{equation}
\label{Tsallis-function}
p(E_1)=\frac{1}{Z_1}\left[1-(q_2-1)\beta^*E_1\right]^{\frac{1}{q_2-1}},
\end{equation}
where $Z_1$ is the partition function, the parameter $\beta^*$ satisfies:
\begin{equation}
\label{beta-star}
\beta^*(q_2-1)=\frac{1}{E_0},
\end{equation}
$E_1$ and $E_0=E_1+E_2$ are the observable and total energies. Equation (\ref{Tsallis-function}) is just another way of writing the distribtuion function of a system in contact with a finite heat bath, therefore, it is not, from the very beginning, a distribution which arises due to the presence of unusual characteristics, such as long-range interactions. All the moments of $E_1$ were calculated in \cite{PoC02}, and equation (\ref{x-relfluct}) was written as:
\begin{equation}
\label{E1-relfluct}
\frac{\left<\Delta E_1^2\right>}{\left<E_1\right>^2}=\frac{q_2/(q_2-1)}{q_1/(q_1-1)}\frac{1}{q_0/(q_0-1)+1},
\end{equation}
where:
\begin{equation}
\label{qs-relation}
\frac{q_0}{q_0-1}=\frac{q_1}{q_1-1}+\frac{q_2}{q_2-1}.
\end{equation}
This expression is nothing less than the additivity of the exponents of the allowed phase space volumes of the three systems.\\
The temperature of a system was defined in \cite{Alm02-01}, through the equipartition theorem, and is given by:
\begin{equation}
\label{temperature}
k_BT=\frac{E_0}{q_0/(q_0-1)}=\frac{\left<E_1\right>}{q_1/(q_1-1)}=\frac{\left<E_2\right>}{q_2/(q_2-1)}
\end{equation}
It may seem confusing the 3 $q$'s introduced above, one for each system, since the whole non-extensive statistics was primarily based on a single parameter $q$. This approach necessarily leads us to the 3 $q$'s, since each system has its own phase space, and therefore, characterized by its own $q$. Fortunately, only one is really important, namely $q_2$. This is the number which characterizes the distribution function of the observable system, and if this $q$ goes to $1$, inother words, if the number of degrees of freedom of the bath goes to infinity, the distribution (\ref{Tsallis-function}) goes to the Boltzmann-gibbs ($BG$) exponential, and the whole formalism, to the classical $BG$ one.\\
The simulation performed here intend to test the validity and accuracy of (\ref{E1-relfluct}). The results are presented in the next section.\\

\section{Numerical models}
\label{section3}
\noindent
The idea here is as follows: suppose we have a system of $N$ particles in $D$ dimensions isolated from any other kind of interaction, which means that it is not coupled to an external bath. It has the following hamiltonian:
\begin{equation}
\label{general-hamiltonian}
H=\sum_{i=1}^{DN}\frac{p_i^2}{2m_i}+U(x_i)+U_{int}(x_i),
\end{equation}
where the first term in the right is the kinetic energy. The other two represent the potential part. Note that there may be an interaction term in this system. Then we pick one degree of freedom, say $p_1$, and separate it from the rest of the hamiltonian:
\[
H=\frac{p_1^2}{2m_1}+\sum_{i=2}^{DN}\frac{p_i^2}{2m_i}+U(x_i)+U_{int}(x_i)=E_1+E_2.
\]
Now we consider $p_1$ as our observable system and the rest of the original hamiltonian as the heat bath. Since the number of degrees of freedom of $E_1$ is finite, in this case equals to $2DN-1$, provided that $DN$ is finite, the distribution of $E_1$, or equivalently $p_1$, is always a power law, and the results found in references \cite{Pros93,Alm02-01,Alm02-02,PoC02-01,Adib02-01,Adib02-02,PoC02} can be tested.\\
Although it seems to be very simple to work in this approach, it should be employed very carefully. Some cases are not direct as they seem to be, we must take into account the mixing of the coordinates.\\
A great example is provided by the three dimensional harmonic oscillator, where $U_{int}(x_i)=0$. In this system we have three momenta and three positions. Hence, it is tempting to use equation (\ref{E1-relfluct}) to $p_1$ considering all the other degrees of freedom as the heat bath. In this case we would have the following numbers:
\[
\frac{q_0}{q_0-1}=3;\frac{q_1}{q_1-1}=\frac{1}{2};\frac{q_2}{q_2-1}=\frac{5}{2}.
\]
These are not correct by the fact that $p_1$ only trades energy with its correspondent position $x_1$, the other coordinates $(p_2,p_3,x_2,x_3)$ do not participate as a bath. In this case, the correct quantities are:
\begin{equation}
\label{q-values}
\frac{q_0}{q_0-1}=1;\frac{q_1}{q_1-1}=\frac{1}{2};\frac{q_2}{q_2-1}=\frac{1}{2}.
\end{equation}
For this reason, we cannot consider the sum of any momenta as a system, for it will not span the whole allowed phase space. Therefore, it is not ergodic.\\
In a system which $U_{int}(x_i)\neq0$, all coordinates trade energy among themselves, and the bath for a single, or more, degree of freedom is defined by the rest of the coordinates. Our second simulation is one of this case. When the system is coupled to an external bath, this carefullness is not needed since this interaction makes all the degrees of freedom to span the whole allowed phase space.\\
This approach was inspired by Khinchin \cite{Khinchin} who suggested that degrees of freedom could be separated as a system and bath.\\
Thus, the simulations we present here do not describe a physical particle. They describes only a degree of freedom of a system.\\

\subsection{The single harmonic oscillator in one dimension}
\label{section3-1}
\noindent
The first system to be simulated is, perhaps, the simplest one to consider. Its hamiltonian reads:
\begin{equation}
\label{osc-hamiltonian}
H=\frac{p^2}{2m}+\frac{1}{2}Kx^2.
\end{equation}
We take $E_1$ as the kinetic part, and the bath as the potential term. The values of $q$ for this problem are given in (\ref{q-values}). We notice that this is a symmetric situation, in the sense that if we choose the potential part as the observable system, the results will be the same. The SRF for this system reads:
\begin{equation}
\label{osc-relfluct}
\frac{\left<\Delta E_1^2\right>}{\left<E_1\right>^2}=\frac{1}{2}.
\end{equation}
The result of the simulations is presented in figure 1. It should be noticed that the SRF does not deviate much from the predicted value.\\
\begin{figure}[ht]
\rotatebox{-90}{\epsfig{file=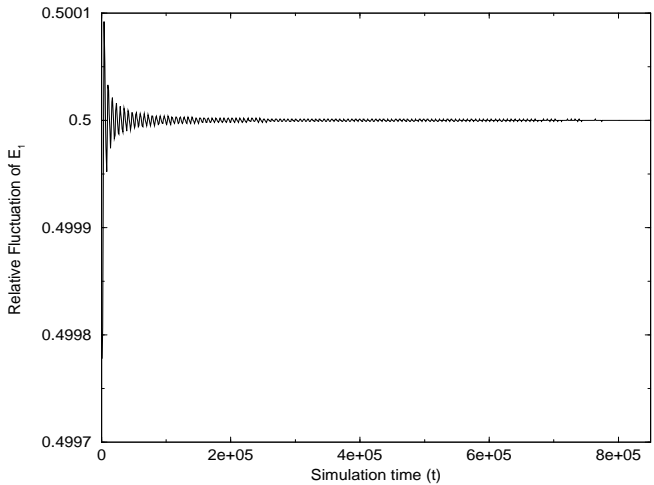,width=6.5cm,height=8.6cm}}
\caption{Relative fluctuation of the kinetic part of the one dimensional oscillator}
\end{figure}
As a by product of the simulation, we present, in figure 3, the fit for the distribution of $E_1$. The slope values $-0.481$. Theoretically predicted value is given by the power of the structure function of the momentum:
\begin{equation}
\label{osc-slope}
\frac{1}{q_2-1}=-0.5.
\end{equation}
\begin{figure}[ht]
\rotatebox{-90}{\epsfig{file=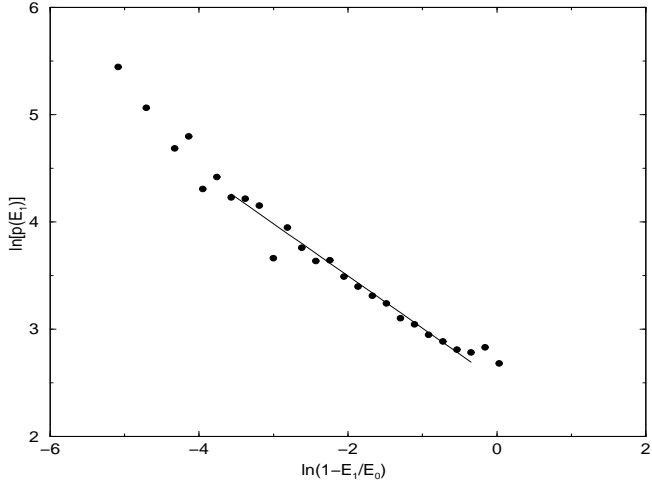,width=6.5cm,height=8.6cm}}
\caption{Linear fit of the distribution of $E_1$}
\end{figure}
In this simulation, we have taken the following values for the important parameters: mass: $m=0.5$, elastic constant: $K=2.0$, time step: $\Delta t=0.01$. The data for building the fit of the distribution were taken after the thermalization time was achieved. We estimated it at about: $t_{ther}=7.0\times10^4$. We did this by observing the time evolution of $Q(t)=\overline{x_i\frac{\partial H}{\partial x_i}}\frac{1}{\Theta}$, where $x_i$ represents the canonical variables, $\Theta$ is proportional to the ratio $\frac{V_0(E_0)}{\Omega_0(E_0)}$, $V_0(E_0)$ is the composed system's accessible volume and $\Omega_0(E_0)$ is its structure function, and is given by:
\[
\Theta=\frac{E_0}{q_0/(q_0-1)},
\]
and the overbar denotes time average. We test the approach of it to unity, for this can be a good estimate of the thermalization time of the system.\\
We used a Leapfrog routine to integrate the system.\\

\subsection{The chain of quartic oscillators}
\label{section3-2}
\noindent
This system was inspired by the Fermi-Pasta-Ulam model, and was first studied in \cite{Adib02-01}. Its hamiltonian is given by:
\begin{equation}
\label{chain-hamiltonian}
H=\sum_{i=1}^{N}\left[\frac{p_i^2}{2}+\frac{x_i^4}{4}+\frac{\left(x_{i+1}-x_i\right)^4}{4}\right],
\end{equation}
where $(p_i,x_i)$ denote the momentum and position of an oscillator, and $N$ is the total number of oscillators.\\
First we will measure the SRF of the kinetic energy of the first oscillator. We choose 
\[
E_1=\frac{p_1^2}{2}; E_2=\sum_{i=2}^{N}\frac{p_i^2}{2}+\sum_{i=1}^{N}\left[\frac{x_i^4}{4}+\frac{\left(x_{i+1}-x_i\right)^4}{4}\right].
\]
From a simple scaling argument \cite{Adib02-01}, we can easily calculate the $q$'s for this problem. They are given next:
\[
\frac{q_0}{q_0-1}=\frac{3N}{4};\frac{q_1}{q_1-1}=\frac{1}{2};\frac{q_2}{q_2-1}=\frac{3N-2}{4}.
\]
Therefore, we have a relative fluctuation given by:
\begin{equation}
\label{chain-relfluct01}
\frac{\left<\Delta E_1^2\right>}{\left<E_1\right>^2}=2\times\frac{3N-2}{3N+4}. 
\end{equation}
In figure 4, we present the results for several values of $N$.\\
\begin{figure}[ht]
\rotatebox{-90}{\epsfig{file=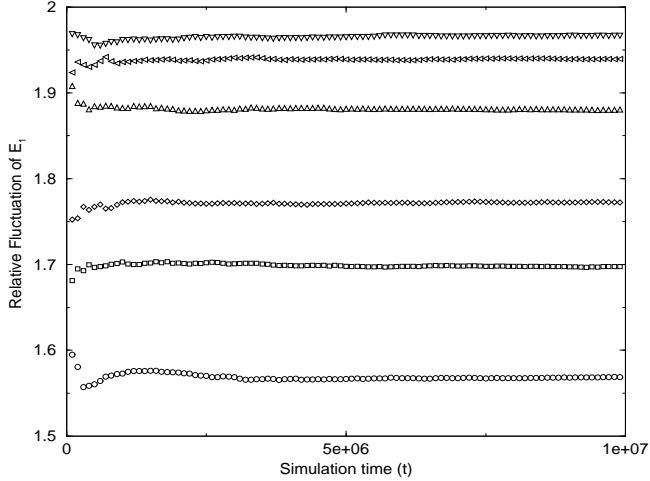,width=6.5cm,height=8.6cm}}
\caption{Relative fluctuation of $\left<p_1^2\right>$ for $N=8,12,16,32,64,128$. The lowest line represents the first case, and the others follow the sequence.}
\end{figure}
In table 1, we show the SRF for the values of $N$ considered in the simulations. We see that while the number of oscillators increases, the SRF approaches $2$, its $N\rightarrow\infty$ value.\\
\begin{table}[ht]
\begin{center}
\begin{tabular}{|c|c|} \hline\hline
$N$  &   $\frac{\left<\Delta E_1^2\right>}{\left<E_1\right>^2}$ \\ \hline 
$08$ &   $1.571$                                                \\ \hline       
$12$ &   $1.7$                                                  \\ \hline
$16$ &   $1.769$                                                \\ \hline
$32$ &   $1.88$                                                 \\ \hline 
$64$ &   $1.938$                                                \\ \hline
$128$ &  $1.969$                                                \\ \hline\hline
\end{tabular}
\caption{Relative fluctuation of $\left<p_1^2\right>$ for several values of $N$}
\end{center}
\end{table}
We also considered a system composed of momenta $p_1, p_2$, only, and then one composed of the first two plus $p_3$, and observed that the SRF agree with (\ref{chain-relfluct01}). This is to be expected since the canonical coordinates are all connected by the quartic interaction.\\
As a second test, we studied the average value of the potential part of this hamiltonian, namely:
\begin{equation}
\label{chain-potential}
E_1 = \sum_{i=1}^{N}\left[\frac{x_i^4}{4}+\frac{\left(x_{i+1}-x_i\right)^4}{4}\right].
\end{equation}
The number of degrees of freedom for this system can be directly calculated by observing that for the kinetic part (which plays the bath's role now), we have:
\[
\frac{q_2}{q_2-1}=\frac{N}{2},
\]
then, we have for the potential energy (observable system):
\begin{equation}
\label{pot-deg-free}
\frac{q_1}{q_1-1}=\frac{N}{4}.
\end{equation}
The SRF, equation (\ref{E1-relfluct}) is given by:
\begin{equation}
\label{pot-relfluct}
\frac{\left<\Delta E_1\right>^2}{\left<E_1\right>^2}=\frac{8}{3N+4}.
\end{equation}
In figure 4, we present the time evolution, after thermalization, of (\ref{pot-relfluct}) for several values of $N$. In table 2, we calculate the values of the relative fluctuation for each case. Once again, if the number of oscillators is increased, the SRF goes to the $N\rightarrow\infty$ value, i. e., zero.\\
\begin{figure}[ht]
\rotatebox{-90}{\epsfig{file=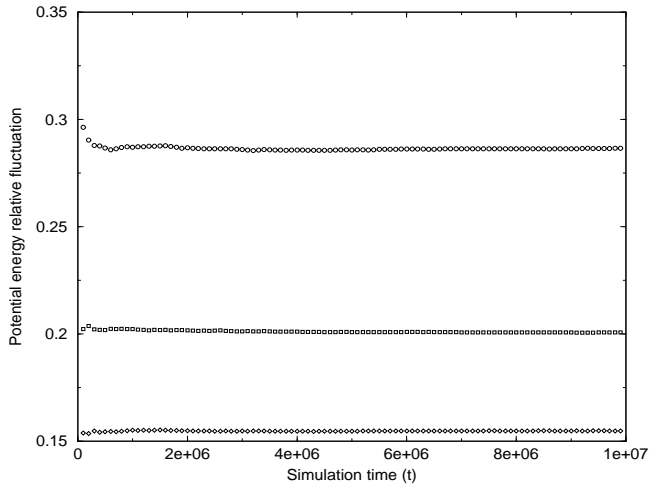,width=6.5cm,height=8.6cm}}
\caption{Relative fluctuation of the potential energy for $N=8,12,16$. The highest line represents the first case, and the others follow the sequence.}
\end{figure}

\begin{table}[ht]
\begin{center}
\begin{tabular}{|c|c|} \hline\hline
$N$ & $\frac{\left<\Delta E_1^2\right>}{\left<E_1\right>^2}$ \\ \hline 
$08$ & $0.285$                                  \\ \hline       
$12$ & $0.2$                                    \\ \hline
$16$ & $0.153$                                          \\ \hline
$32$ & $0.08$                                           \\ \hline 
$64$ & $0.04$                                           \\ \hline
$128$ & $0.02$                                          \\ \hline\hline
\end{tabular}
\caption{Relative fluctuation of the mean potential energy for several values of $N$}
\end{center}
\end{table}
A third test of relation (\ref{E1-relfluct}) would be to study only a term in the potential energy, say $x_0^4$. The $q_1/(q_1-1)$ values may be, in some cases, not so hard to calculate, and (\ref{E1-relfluct}) would be readly applicable. Unfortunately, this procedure does not work, as we observed by studying the above term. Although its mean value and SRF converge, these values are not predicted by our calculations. We observed that this term does not obeys the equipartition, in other words, its mean value does not follow:
\[
\left<x_0^4\right>=\frac{1}{4}k_BT.
\]
This suggests that isolated terms in the potential are non-ergodic, although the whole potential energy is an ergodic one.\\
In figure 5, we show the time evolution of the quantity:
\[
\frac{\overline{x_0^4}}{4k_BT}
\]
which, by the former, should approach $0.25$. Clearly, we see that it does not. We have also observed this behaviour in other terms, such as $x_1^4-x_0^4$.\\
Although the equipartition does not hold in this case, we could use the final value of $q_1/(q_1-1)$ and calculate the relative fluctuation of this term. Once again, the value obtained does not coincide with the one observed in the simulations, not shown here. Hence, we cannot predict, with the tools of the canonical ensemble, for this $x_0^4$ term, any quantity, and, by the nature of the problem, we can also say that the other terms in the potential energy follow this line.\\
The reason for this discrepancy may lie in the fact that no term in the potential energy is independent of any other term in the potential energy. In the kinetic part, no single momentum component is free, it interacts with the potential part, but it does not interact with the other momenta components. This is why the single terms in the potential energy do not span the allowed phase space.\\
\begin{figure}[ht]
\rotatebox{-90}{\epsfig{file=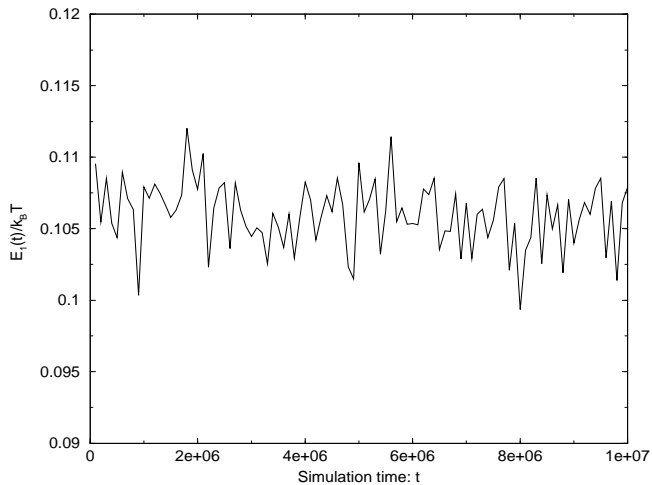,width=6.5cm,height=8.6cm}}
\caption{Time evolution of $\frac{\overline{x_0^4}}{k_BT}$ for $N=8$.}
\end{figure}

\section{Discussion}
\label{section4}
\noindent
We saw that our separation of degrees of freedom scheme provides a simple way of studying systems with a finite number of degrees of freedom. The results obtained here are in very good agreement with the theoretical results. This fact indicates that the time averages of such finite systems coincide with their ensemble averages. This seems to be in contradiction with the thought that for statistical mechanics to apply, it is required infinite number of particles, or infinite number of degrees of freedom. We state that this represents no contradiction at all for ensemble averages are taken with respect to the system's states of energy, and not with respect to the system's degrees of freedom. The number of states of energy of a system are always infinite in number for a classical system (which is the case here).\\
For example, consider a classical particle in a box. How many states of energy, i.e. how many values of the canonical coordinates $(\vec p, \vec x)$, exist such that are compatible with the particle's energy? Clearly infinite. Generally, every degree of freedom defines a surface in its own phase space, with an infinite number of points (states).\\

\section{Conclusions}
\label{section5}
\noindent
We have simulated two simple systems not in thermal equilibrium with a heat bath, in order to separate its degrees of freedom, and consider them as separate systems, one as the observable and the other as the heat bath. We intended to study systems with finite number of degrees of freedom, and therefore, to test the results found in \cite{PoC02} for the square relative fluctuation of the observable system energy.\\
Our results are in very good agreement with the theoretical predicted value, hence indicating that these systems are indeed ergodic.\\
We can conclude that these finite systems do not represent a case of ``bad'' statistics, in the sense that the methods of statistical mechanics do not apply to them. Certainly, the method of Boltzmann-Gibbs, which assumes an ideal thermostat in certainly not appplicable, but our procedure of considering the system decoupled from the ideal heat bath, hence not in equilibrium with such a system, is clearly a good framework to use.\\

\end{document}